\documentclass[aps,amssymb,12pt,floatfix]{revtex4}
 \usepackage{subfigure}
\usepackage{graphicx} \usepackage{rotating}

\begin{document}

\baselineskip=14pt

\title{Mapping the energy landscape of biomolecules using single molecule
force correlation spectroscopy (FCS): Theory and applications}

\author{V. Barsegov$^1$, D. K. Klimov$^3$  and
D. Thirumalai$^{1,2}$} \thanks{Corresponding author phone:
301-405-4803; fax: 301-314-9404;  thirum@glue.umd.edu}
\affiliation{$^1$Biophysics Program, Institute for Physical Science and Technology,
$^2$Department of Chemistry and Biochemistry\\ University of Maryland,
College Park, MD 20742\\ $^3$Bioinformatics and Computational Biology Program,
School of Computational Sciences\\ George Mason University, Manassas, VA 20110}

\date{\today}
\begin{abstract}

\baselineskip=14pt

We present a new theory that takes internal dynamics of proteins into account to describe
forced-unfolding and force-quench refolding in single molecule experiments. In the current 
experimental setup (Atomic Force Microscopy or Laser Optical Tweezers) the distribution of 
unfolding times, $P(t)$, is measured by applying a constant stretching force ${\bf f}_{S}$ 
from which the apparent ${\bf f}_{S}$-dependent unfolding rate is obtained. To describe the complexity 
of the underlying energy landscape requires additional probes that can incorporate the dynamics of 
tension propagation and relaxation of the polypeptide chain upon force quench. We introduce a theory 
of force correlation spectroscopy (FCS) to map the parameters of the energy landscape of proteins. 
In the FCS the joint distribution $P(T,t)$ of folding and unfolding times is constructed by repeated 
application of cycles of stretching at constant ${\bf f}_{S}$ separated by release periods $T$ during
which the force is quenched to ${\bf f}_{Q}$$<$${\bf f}_{S}$. During the release period, the protein 
can collapse to a manifold of compact states or refold. We show that $P(T,t)$ at various ${\bf f}_{S}$ 
and ${\bf f}_{Q}$ values can be used to resolve the kinetics of unfolding as well as formation of 
native contacts. We also present methods to extract the parameters of the energy landscape using 
chain extension as the reaction coordinate and $P(T,t)$. The theory and a worm-like chain model for 
the unfolded states allows us to obtain the persistence length $l_p$ and the ${\bf f}_{Q}$-dependent 
relaxation time, that gives an estimate of collapse timescale at the single molecular level, in the 
coil states of the polypeptide chain. Thus, a more complete description of landscape of protein  
native interactions can be maped out if unfolding time data are collected at several  values of 
${\bf f}_{S}$ and ${\bf f}_{Q}$. We illustrate the utility of the proposed formalism by analyzing 
simulations of unfolding-refolding  trajectories of a coarse-grained protein ($S1$) with $\beta$-sheet 
architecture for several values of  ${\bf f}_{S}$, $T$ and ${\bf f}_{Q}$$=$$0$. The simulations of 
stretch-relax trajectories are used to map many of the parameters that characterize the energy 
landscape of $S1$.

\end{abstract}
\maketitle

\section{\bf INTRODUCTION}

Several biological functions are triggered by mechanical force. These include stretching
and contraction of muscle proteins such as titin  \cite{1,2}, rolling and tethering of
cell adhesion molecules \cite{3,4,4new,4a,4b}, translocation of proteins across membranes \cite{5,5a,6,6a},
and unfoldase activity of chaperonins and proteasomes. Understanding these diverse functions 
requires probing the response of biomolecules to applied external tension. Dynamical responses 
to mechanical force can be used to characterize in detail the free energy landscape of biomolecules. 
Advances in manipulating micron-sized  beads attached to single biomolecules have made it possible to 
stretch, twist, unfold and even unbind proteins using forces on the order of tens of piconewtons 
\cite{7,7a,7b}. Single molecule force spectroscopy on a number of different systems has allowed us to 
obtain a glimpse of the unbinding energy landscape of biomolecules and protein-protein complexes
\cite{8,9,10,10a}. In AFM experiments, used to unfold proteins by force, one end of 
a protein is adsorbed  on a template and a constant or a time-dependent pulling force is applied 
to the other terminus \cite{11,12,12a,12b,13,13a,14}. By measuring the distribution of forces required 
to completely unfold proteins and the associated unfolding times, the global parameters of the protein 
energy landscape can be estimated \cite{15,16,17,18,18a,19}. These insightful experiments when 
combined with theoretical studies \cite{19a,19b,19c} can give an unprecedented picture of 
forced-unfolding pathways.

Current experiments have been designed primarily to obtain information on forced-unfolding 
of proteins and do not  probe the reverse folding process. Although force-clamp AFM techniques have 
been used recently to probe (re)folding of single ubiquitin polyprotein \cite{13a}, the lack of 
theoretical approaches has made it difficult to interpret these pioneering experiments \cite{19e,19f}. 
Secondly, the resolution of multiple timescales in protein folding and refolding requires not only 
novel experimental tools for single molecule experiments but also new theoretical analysis methods. 
Minimally, unfolding of proteins by a stretching force ${\bf f}_{S}$ is described by the global 
unfolding time $\tau_U({\bf f}_{S})$, timescales for propagation of the applied tension, and the 
dynamics describing the intermediates or ``protein coil'' states. Finally, if the external conditions 
(loading rate or the magnitude of ${\bf f}_{S}$) are such that these processes can occur on similar 
timescales then the analysis of the data requires new theoretical ideas. 

For forced unfolding the variable conjugate to ${\bf f}_{S}$, namely, the protein 
end-to-end distance $X$ is a natural reaction coordinate. However, $X$ is not appropriate for 
describing protein refolding which, due to substantial variations in the duration of
folding  barrier crossing, may range from milliseconds to few minutes. To obtain statistically 
{\it meaningful} distributions of unfolding times, a large  number of {\it complete} unfolding 
trajectories must be recorded which requires repeated application of the pulling force. 
The inherent heterogeneity in the duration of folding and the lack of correlation between 
evolution of $X$ and (re)folding progress creates ``initial state ambiguity'' when force is repeatedly 
applied to the same molecule. As a result, the interpretation of unfolding time data is complicated
especially when the conditions are such that reverse folding process at the quenched force ${\bf f}_{Q}$ 
can occur on a long timescale, $\tau_F({\bf f}_{Q})$.

Motivated by the need to assess the effect of the multiple timescales on the energy landscape
of folding and unfolding, we develop a new theoretical formalism to describe correlations 
between the various dynamical processes. Our theory leads naturally to a new class of
single molecule force experiments, namely, the force correlation spectroscopy (FCS) which
can be used to study both forced unfolding as well as force-quenched (re)folding. Such studies
can lead to a more detailed information on both kinetic and dynamic events underlying unfolding and 
refolding. In the FCS, cycles of stretching (${\bf f}_{S}$) are separated by periods $T$ of quenched 
force  ${\bf f}_{Q}$$<$${\bf f}_{S}$ during which the stretched protein can relax 
from its unfolded state $X_U$ to coil state $X_C$ or even (re)fold to the native basin of 
attraction (NBA) state. The two experimental observables are $X$ and the unfolding time $t$. 
The central quantity in the FCS is the distribution of unfolding times $P(T,t)$ separated by recoil or 
refolding events of duration $T$. The higher order statistical measure embedded in $P(T,t)$ is readily 
accessible by constructing a histogram of unfolding times for varying $T$ and does not require additional 
technical developments. The crucial element in the proposed analysis is that $P(T,t)$ is computed by 
{\em averaging over final (unfolded) states}, rather than initial (folded) states. This procedure 
removes the potential ambiguity of not precisely knowing the initial distribution of conformations 
in the NBA. Despite the uniqueness of the native state there are a number of conformations in the NBA
that reflect the fluctuations of the folded state. The proposed formalism is a natural extension of 
unbinding time data analysis. Indeed, $P(T,t)$ reduces to the standard distribution of unfolding times 
$P(t)$ when $T$ exceeds protein (re)folding timescale $\tau_F({\bf f}_{Q})$.

The complexity of the energy ladscape of proteins demands FCS and the theoretical analysis.
Current single molecule experiments on poly-Ub or poly-Ig27 (performed in the $T$$\to$$\infty$ regime)
show that in these systems unfolding occurs abruptly in an apparent all-or-none manner or through a 
dominant intermediate \cite{19a}. On the other hand, refolding upon force-quench is complex and surely 
occurs though an ensemble of collapsed coiled states \cite{13a}. A number of timescales characterize the 
stretch-release experiments. These include besides $\tau_F({\bf f}_{Q})$, the ${\bf f}_{S}$-dependent 
unfolding time, and the relaxation dynamics in the coiled states $\{ C \}$ upon force-quench 
$\tau_d({\bf f}_{Q})$. In addition, if we assume that $X$ is an appropriate reaction coordinate then 
the location of the NBA, $\{ C \}$, the transition state ensembles and the associated widths are required 
for a complete characterization of the underlying energy landscape. Most of these parameters can be 
extracted using the proposed FCS experiments and the theoretical analysis presented here.

In a preliminary study \cite{PRL}, we reported the basics of the theory which was used to propose
a new class of single molecule force spectroscopy methods for deciphering protein-protein interactions.
The current paper is devoted to further developments in the theory with appplication to forced-unfolding
and force-quench refolding of proteins. In particular, we illustrate the efficacy of the FCS by analyzing 
single unfolding-refolding trajectories generated for a coarse-grained model (CGM) protein $S1$ with 
$\beta$-sheet architecture \cite{19new,20old}. We showed previously that forced-unraveling of $S1$, in the 
limit of $T$$\to$$\infty$, can be described by an apparent ``two-state'' kinetics \cite{20,21}. 
The thermodynamics and kinetics observed in $S1$ is a characteristic of a number of proteins where
folding/unfolding fits well two-state behavior \cite{21new}. Thus, $S1$ serves as a useful model to 
illustrate the efficacy of the FCS. Here, we show that by varying $T$ and the magnitude of the 
stretching (${\bf f}_{S}$ or ${\bf f}_{Q}$), the entire dynamical processes, starting from the NBA to 
the fully stretched state, can be resolved. In the process we establish that $P(T,t)$ which can be measured 
using AFM or laser optical tweezer (LOT) experiments, provides a convenient way of characterizing the energy 
landscape of biomolecules in detail.

\section{\bf Models and Methods:}

{\it Theory of force correlation spectroscopy (FCS):} In single molecule atomic force microscopy (AFM) 
experiments used to unfold proteins by force, the N-terminus of a  protein is anchored at the surface 
and the C-terminus is attached to the cantilever tip  through a polymer linker (Figure 1). The molecule 
is stretched by displacing the cantilever tip and the resulting force is measured. From a theoretical 
perspective it is more convenient to envision applying a constant stretching force 
${\bf f}_{S}$$=$$f_S$${\bf x}$ in the ${\bf x}$-direction (Figure 1). The free energy in the constant 
force formulation is related to the experimental setup by a Legendre transformation. More recently, it 
has become possible to apply a constant force in AFM or laser or optical tweezer (LOT) experiments to 
the ends of a protein. With this setup the unfolding time for the end-to-end distance $X$ to reach the 
contour length $L$ can be measured for each molecule. For a fixed  ${\bf f}_{S}$, repeated application 
of the pulling force results in a single trajectory of  unfolding times ($t_1$, $t_2$, $t_3$, $\ldots$, 
Figure 1) from which the histogram of unfolding times $P(t)$ is obtained. The ${\bf f}_{S}$-dependent 
unfolding rate $K_U$ is obtained by fitting a Poissonian formula $K_U^{-1}\exp{[-K_U t]}$ to the kinetics 
of population of folded states $p_F$ which is related to $P(t)$ as $p_F(t)$$=$$1$$-$$\int_0^t ds P(s)$. 

Because $K_U$ is a convolution of several microscopic processes, it does not describe unfolding
in molecular detail. For instance, mechanical unfolding of fibronectin domains FnIII involves the 
intermediate ``aligned'' state \cite{16} with partially disrupted hydrophobic core which cannot be 
resolved by knowing only $K_U$. Even when the transition from the folded state $F$ to the globally 
extended state $U$ \cite{16} does not involve parallel routes as in Figure 2, or multistate kinetics,
the force-induced unfolding pathway must involve formation of intermediate coiled states $\{ C \}$. 
The subsequent transition from $\{ C \}$ results in the formation of the globally unfolded state $U$. 
The incomplete time resolution prevents current experiments from probing the signature of the collapsed 
states. To probe the contributions from the underlying $\{ C \}$ states to global unfolding requires 
sophisticated experiments that can resolve contributions from dynamic events underlying forced unfolding. 
We propose a novel experimental procedure which, when supplemented with unfolding time data analysis 
described below, allows us to separately probe the kinetics of native interactions and the dynamics of the 
``protein coil'' (i.e. the dynamics of end-to-end distance $X$ when the native contacts are disrupted).

Consider an experiment in which stretching cycles (triggered by applying ${\bf f}_{S}$) are interrupted 
by relaxation intervals $T$ during which force is quenched to ${\bf f}_{Q}$$<$${\bf f}_{S}$. In the 
time interval $T$, the polypeptide chain can relax into the manifold $\{ C \}$ or even refold to the 
native state $F$ if $T$ is long enough. If ${\bf f}_{S}$$>$${\bf f}_C$ and ${\bf f}_{Q}$$<$${\bf f}_C$ 
where ${\bf f}_C$ is the equilibrium critical unfolding force at the specific temperature (see phase 
diagram for $S1$ in Ref.~\cite{20}), these transformations can be controlled by $T$. In the simplest 
implementation we set ${\bf f}_{Q}$$=$$0$. The crucial element in the FCS experiment is that the same 
measurements are repeated for varying $T$. In the FCS the unfolding times are binned to obtain the 
joint histogram $P(T,t)$ of unfolding events of duration $t$ generated from the recoil manifold $\{ C\}$ 
or the native basin of attraction (NBA) or both, depending on the duration of the relaxation time $T$. 
In the current experiments $T$$\to$$\infty$. As a result, the dynamics of additional states in the 
energy landscape that are explored during folding or unfolding are not probed.

The advantages of $P(T,t)$ over the standard distribution  of unfolding times $P(t)$ are two-fold. First, 
$P(T,t)$ is computed by {\em averaging over well-characterized fully stretched states}. This eliminates 
the problem of not knowing the distribution of initial protein states encountered in current experiments. 
Indeed, due to intrinsic heterogeneity of the protein folding pathways, after the first unfolding event 
the protein may or may not refold into the native conformation, which creates the initial state ambiguity 
in the next (second, third, etc) pulling cycle. Therefore, statistical analysis based on averaging over 
final (stretched) states rather than initial (folded) states allows to overcome this difficulty. Secondly, 
statistical analysis of unfolding data {\em performed for different values of $T$} allows us to 
separately probe the kinetics of native interactions and the dynamics of $X$. In addition, the entire 
energy landscape of native interactions can be mapped out when stretch-quench cycles are repeated for 
several values of ${\bf f}_{S}$, ${\bf f}_{Q}$, and $T$.

{\it Regime I} ($T$$\ll$$\tau_F$): In the simplest unfolding scenario application 
of ${\bf f}_{S}$ results in the disruption of the native contacts ($F$$\to$$\{ C \}$) followed 
by stretching of the manifold $\{ C \}$ into $U$ (Figure 2). When stretching cycles are 
separated by short $T$ compared to the protein folding timescale  $\tau_F$ at ${\bf f}_Q$$=$$0$, 
$P(T;t)$ is determined by the evolution of the coil state. Then the unfolded state population $p_U(T;t)$ 
is given by the convolution of protein relaxation (over time $T$) from the fully stretched state 
$X_U$$\approx$$L$ to an intermediate coiled state $X_1$ and streching $X_1$ into final state $X_f$ over 
time $t$. Thus, $P(T;t)$ is obtained from $p_U(T;t)$ by taking the derivative with respect to $t$,
\begin{eqnarray}\label{2.1}
P(T\ll \tau_F;t) & = & {{d}\over {dt}}p_U(T\ll \tau_F;t)\\\nonumber &
                 = & {{d}\over {dt}}{{1}\over {N(T)}}\int_{L-\delta}^L
                 dX_f 4\pi X_f^2  \int_0^L dX_1 4\pi X_1^2 \int_0^L
                 dX_U 4\pi X_U^2\\\nonumber  & \times &
                 G_S(X_f,t;X_1)G_Q(X_1,T;X_U)P(X_U)
\end{eqnarray}
where $N(T)$ is $T$-dependent normalization constant obtained by taking the last integral in the 
right hand side (rhs) of Eq. (\ref{2.1}) from $X_f$$=$$0$ to $X_f$$=$$L$, and $P(X_U)$ is the 
distribution of unfolded states. If $X$ is well controlled, $X_U$ is expected to be centered around 
a fixed value ${\bar X}_U$ and $P(X_U)$$\sim$$\delta(X_U-{\bar X}_U)$. In Eq. (\ref{2.1}), 
$G_Q(X',t;X)$ and $G_S(X',t;X)$ are respectively, the quenched and the stretching force 
dependent conditional probabilities to be in the coiled state $X'$ at time $t$ arriving 
from state $X$ at time $t$$=$$0$. The integral over $X_f$ is performed in the range 
$[ L-\delta;L ]$ with  $X$$=$$L$$-$$\delta$ (Figure 2) representing unfolding distance at 
which the total number of native contacts $Q$ is at the unfolding threshold, $Q$$\approx$$Q^*$. 
It follows that $P(T;t)$ (Eq. (\ref{2.1})) contains information on the dynamics of $X$. 
By assuming a model for $X$ and fitting $P(T;t)$, obtained by differentiating the integral expression
appearing in Eq. (\ref{2.1}), to the histogram of unfolding times, separated by short $T$$\ll$$\tau_F$, 
we can resolve the dynamics of the polypeptide chain in the coil state which allows us to evaluate 
the ${\bf f}_Q$-dependent coil dynamical timescale $\tau_d$ using single molecule force spectroscopy.
The fit of Eq. (\ref{2.1}) could be analytical or numerical depending on the model of $X$.

{\it Regime II} ($T$$\gg$$\tau_F$): When stretching cycles are interrupted by long relaxation periods, 
$T$$\gg$$\tau_F$, the coiled states refold to $X_F$ (Figure 2). In this regime, the initial conformations 
in forced-unfolding always reside in the NBA. In this limit, $P(T;t)$ reduces to the standard distribution 
of unfolding times $P(T,t)$$\to$$P(t)$. When $T$$\gg$$\tau_F$, $P(T;t)$ is given by the convolution of the 
kinetics of rupture of native contacts, resulting in protein extension $\Delta X_F$, and dynamics of $X$ 
from state $X_F$$+$$\Delta X_F$ to final state $X_f$, 
\begin{eqnarray}\label{2.2}
P(T\gg\tau_F;t) & = & P(t)={{d}\over {dt}}p_U(T\gg\tau_F;t)\\\nonumber
                & = & {{d}\over {dt}}{{1}\over
                {N'(T)}}\int_{L-\delta}^L dX_f 4\pi X_f^2  \int_0^L
                dX_F 4\pi X_F^2 \int_0^t dt'\\\nonumber & \times &
                G_S(X_f,t;X_F+\Delta X_F,t') P_F(t',X_F;{\bf f}_{S})
\end{eqnarray}
where $N'(T)$ is normalization constant obtained as in Eq. (\ref{2.1}) and $P_F(t,X_F;{\bf f}_{S})$ is 
the probability of breaking the contacts over time $t$ that stabilize the native state $X_F$. By assuming a 
model for $P_F(t,X_F;{\bf f}_{S})$ and employing information on the dynamics of $X$, obtained from the short 
$T$-experiment (Eq. (\ref{2.1})), we can probe the disruption kinetics of native interactions. By repeating 
long $T$-measurements at several values of ${\bf f}_{S}$, we can map out the energy landscape of native 
interactions projected on the direction of the end-to-end distance vector. 

{\it Regime III} ($T$$\sim$$\tau_F$): In this limit, some of the molecules reach the NBA, starting 
from extended states ($X$$\approx$$L$), whereas others remain in the basin $\{ C \}$. The fraction of folding 
events $\rho_F$ depends on $T$ during which $X$ approaches the average extension $\langle X_C \rangle$ 
facilitating the formation of native contacts. Thus, $P(T\sim \tau_F)$ obtained in the intermediate 
$T$-experiment, involves contributions from both $\{ C \}$ and $F$ initial conditions and is given by a 
superposition,
\begin{equation}\label{2.4}
P(T\sim \tau_F;t)=\rho_F(T)P(T\gg\tau_F;t) + \rho_C(T)P(T\ll \tau_F;t)
\end{equation}
where the probability to arrive to $F$ from $\{ C \}$ at time $T$ is given by
\begin{equation}\label{2.5}
\rho_F(T)=\int_0^LdX_1 4\pi X_1^2 \int_0^L dX_U 4\pi X_U^2 P_C(T, X;{\bf f}_{Q}) G_Q(X_1,T;X_U)P(X_U)
\end{equation}
and the probability to remain in $\{ C \}$ is $\rho_C(T)$$=$$1$$-$$\rho_F(T)$. 
In Eq. (\ref{2.5}), $P_C(T, X;{\bf f}_{Q})$ is the refolding probability 
determined by kinetics of formation of native contacts. Because the dynamics of 
$X$ is weakly correlated with formation of native contacts, $X$ in $P_C$ is expected
to be broadly distributed. Therefore, Eqs. (\ref{2.4}) and (\ref{2.5}) can be used
to probe kinetics of formation of native interactions.

For Eqs. (\ref{2.1}) and (\ref{2.2}) to be of use, one needs to know the (re)folding timescale 
$\tau_F$. The simplest way to evaluate $\tau_F$ is to  construct a series of histograms 
$P(T_n,t)$ ($n$$=$$1$, $2$, $\ldots$, $N$) for a fixed  ${\bf f}_{S}$ and increasing relaxation time 
$T_1$$<$$T_2$$<$$\ldots$$<$$T_N$, and compare $P(T_n,t)$'s with the distribution $P(T^*,t)$ 
obtained for sufficiently long $T^*$$\gg$$\tau_F$. If $T$$=$$T^*$ then all the molecules are
guaranteed to reach the NBA. The difference
\begin{equation}\label{2.3}
D(T_n) = |P(T_n,t)-P(T^*,t)|
\end{equation}
is expected to be non-zero for $T_n$$\le$$\tau_F$ and should vanish if $T_n$ exceeds $\tau_F$. 
Statistically, as $T_n$ starts to exceed $\tau_F$ increasingly more molecules will reach the NBA 
by forming native contacts. Then, more unfolding trajectories will start from folded states, and 
when  $T$$\gg$$\tau_F$ all unfolding events will originate from the NBA. Therefore, $D(T_n)$ is a 
sensitive measure for identifying the kinetic signatures for forming native contacts. The utility
of $D(T_n)$ is that it is a simple yet accurate estimator of $\tau_F$, which can be utilized in 
practical applications. Indeed, one can estimate $\tau_F$ by identifying it with the shortest $T_n$ 
at which $P(T_n;t)$$\approx$$P(T^*,t)$, i.e. $T_n$$\approx$$\tau_F$. We should emphasize that to obtain
$\tau_F$ from the criterion that $D(\tau_F)$$\approx$$0$ no assumption about the distribution of 
refolding times have been made. Having evaluated $\tau_F$ one can then use Eqs. (\ref{2.1}) and 
(\ref{2.2}) for short and long $T$-measurements to resolve protein coil dynamics and rupture kinetics 
of native contacts.

Let us summarize the major steps in the FCS. First, we estimate $\tau_F$ by using $D(T)$ (Eq. \ref{2.3})). 
We next probe protein coil dynamics by analyzing $P(T\ll \tau_F;t)$ obtained from short-$T$-measurements 
(Eq. (\ref{2.1})). In the third step, we use information on protein coil dynamics to resolve the kinetics 
of rupture of native interactions contained in $P(T\gg\tau_F;t)$ of long-$T$-measurements 
(Eq. (\ref{2.2})). Finally, by employing the information on protein coil dynamics and kinetics of rupture 
of native interactions, we resolve the kinetics of formation of native contacts by analyzing 
$P(T\sim \tau_F;t)$ from intermediate $T$-measurements (Eqs. (\ref{2.4}) and (\ref{2.5})). 

The beauty of the proposed framework is that these experiments can be readily performed using available 
technology. In the current AFM experiments, $T$ can be made as short as few microseconds. Simple calculations
show that the relaxation of a short $50$ amino acid protein from stretched state with $L$$\approx$$19nm$ 
to the coiled states $\{ C \}$ with say, $X$$\approx$$2nm$, occurs on the timescale 
$\tau_d$$\approx$$\Delta x^2/D$$\sim$$10\mu s$, where $\Delta x$$=$$L$$-$$X$$\approx$$17nm$ and 
$D$$\approx$$10^{-7}cm^2/s$ is the diffusion constant. Clearly, the time of formation of native
contacts, which drives the transition from $\{ C \}$ to the NBA, prolongs $\tau_F$ by few microseconds
to few miliseconds or larger, depending on folding conditions. In the experimental studies of forced 
unfolding and force-quenched refolding of ubiquitin, $\tau_F$ was found to be of the order of $10$$-$$100ms$ 
\cite{13a}. Computer simulation studies of unzipping-rezipping transitions in short $22$-nt RNA hairpin 
P5GA have predicted that $\tau_F$ is of the order of few hundreds of microseconds \cite{19e}.

{\it Model for the kinetics of native contacts}: To interpret the data generated by FCS it is useful to have
a model for the time evolution of the native contacts and $X$. We first present a simple kinetic model for 
rupture and formation of native contacts represented by probabilities $P_F$ and $P_C$ in Eqs. (\ref{2.2}) and 
(\ref{2.5}), respectively, and a model for the dynamics of $X$ given by the propagator $G_{S,Q}(X',t;X)$. 
To describe the force-dependent evolution of native interactions we adopt the continuous-time-random-walk 
(CTRW) formalism \cite{22,22a,24,24a,24b}. In the CTRW model, a random walker, representing rupture 
(formation) of native contacts, pauses in the native (coiled) state for a time $t$ before making a 
transition to the coiled (native) state. The waiting time distribution is given by the function 
$\Psi_{\alpha}(t)$ ($\alpha$$=$$r$ or $f$, where $r$ and $f$ refer to rupture and formation of native contacts,
respectively). We assume that the probabilities $P_F(t,X_F;{\bf f}_{S})$ and $P_C(t,X_C;{\bf f}_{Q})$ are 
separable so that 
\begin{equation}\label{3.2}
P_F(t,X_F;{\bf f}_{S})\approx P_{eq}(X_F)P_r(t;{\bf f}_{S}),\quad
\text{and} \quad  P_C(t,X_C;{\bf f}_{Q})\approx P_C(X_C)P_f(t;{\bf f}_{Q})
\end{equation}
where $P_{eq}(X_F)$ is the equilibrium distribution of native states, $P_C(X_C)$ is the distribution of 
coiled states and $P_r(t;{\bf f}_{S})$ and $P_f(t;{\bf f}_{Q})$ are the force-dependent probabilities of 
rupture and formation of native contacts, respectively. Factorization in Eq. (\ref{3.2}) implies that 
application of force does not result in the redistribution of states $X_F$ and $X_C$ in the NBA and in the 
manifold of coiled states $\{ C \}$, but only changes the timescales for NBA$\to$$\{ C \}$ and 
$\{ C \}$$\to$NBA transitions, and thus, the propabilities $P_r$ and $P_f$. We expect the approximation in 
Eq. (\ref{3.2}) to be valid provided the rupture of native contacts and refolding events are cooperative.

During stretching cycles, for ${\bf f}_{S}$ well above ${\bf f}_C$, we may neglect the reverse 
folding process. Similarly, global unfolding is negligible during relaxation periods with 
${\bf f}_{Q}$$<$${\bf f}_C$. Then, the master equations for $P_r(t)$ is 
\begin{equation}\label{3.3}
{{d}\over {dt}}P_{r}(t)= -\int_0^t d\tau \Phi_{r}(\tau)P_{r}(t-\tau)
\end{equation}
where $\Phi_{r}(t)$ is the generalized rate for the rupture and formation of native interactions. 
In the Laplace domain, defined by ${\bar f}(z)$$=$$\int_0^{\infty}dt f(t)\exp{[-tz]}$, $\Psi_r(t)$ 
is related to $\Phi_{r}(t)$ as
\begin{equation}\label{3.4}
{\bar \Phi}_{r}(z)=z{\bar \Psi}_{r}(z)\left[ 1-{\bar \Psi}_{r}(z)
\right]^{-1}.
\end{equation}
The structure of the master equation for $P_f(t)$ is identical to Eq. (\ref{3.3}) with the relationship
between $\Phi_{f}(t)$ and $\Psi_{f}(t)$ being similar to Eq. (\ref{3.4}). The general solution to 
Eqs. (\ref{3.3}) is
\begin{equation}\label{3.5}
{\bar P}_{r}(z)=\left[ z-\Phi_{r}(z)\right]^{-1}{\bar P}_{r}(0)
\end{equation}
where ${\bar P}_{r}(0)$$=$$1$ is the initial condition and the solution in the time domain is given
by the inverse Laplace transform $P_{r}(t)$$=$$L^{-1}\{ {\bar P}_{r}(z) \}$. The solution
for ${\bar P}_{f}(z)$ is obtained in a similar fashion (see Eq. (\ref{3.5})) with initial 
condition ${\bar P}_{f}(0)$$=$$1$.

{\it Model for the polypeptide chain}: In the extended state, when the majority of native interactions 
that stabilize the folded state are disrupted, the molecule can be treated roughly as a fluctuating coil. 
Simulations and analysis of native structures \cite{19new} suggest that proteins behave as 
worm-like chains (WLC). For convenience we use a continuous WLC description for the coil 
state whose Hamiltonian is
\begin{eqnarray}\label{4.1}
H & = & {{3k_B T}\over {2l_p}}\int_{-L/2}^{L/2} ds \left({{\partial
{\bf r}(s,t)} \over {\partial s}}\right)^2 + {{3l_p k_B T}\over
{8}}\int_{-L/2}^{L/2} ds\left({{\partial^2 {\bf r}(s,t)} \over
{\partial s^2}}\right)^2  \\\nonumber & + & {{3 k_b T}\over {4}}\left[
\left({{\partial {\bf r}(-L/2,t)} \over {\partial s}}\right)^2 +
\left({{\partial {\bf r}(-L/2,t)}\over {\partial s}}\right)^2 \right]+
{\bf f}_{S,Q}\int_{-L/2}^{L/2}ds \left({{\partial {\bf r}(s,t)} \over
{\partial s}}\right)
\end{eqnarray}
where $l_p$ is the protein coil persistence length. A large number of force-extension curves obtained
using mechanical unfolding experiments in proteins, DNA, and RNA have been analyzed using WLC model.
In Eq. (\ref{4.1}) the three-dimensional Cartesian vector ${\bf r}(s,t)$ represents the spatial location 
of the  $s^{th}$ ``protein monomer'' at time $t$. The first two terms describe chain connectivity and 
bending energy, respectively. The third term represents fluctuations of the chain free ends and the 
fourth term corresponds to coupling of ${\bf r}$ to ${\bf f}_{S,Q}$. The end-to-end vector is computed 
as ${\bf X}(t)$$=$${\bf r}(L/2,t)$$-$${\bf r}(-L/2,t)$.

We need a dynamical model in which $X$ is represented by the propagator $G(X,t;X_0)$.
Although bond vectors of a WLC chain are correlated, the statistics of $X$  can be 
represented by a large number of independent modes. It is therefore reasonable, at
least in the large $L$ limit, to describe $G_{S,Q}(X,t;X_0)$ by a Gaussian, 
\begin{equation}\label{4.2}
G_{S,Q}(X,t;X_0) = \left( {{3}\over {2 \pi \langle X^2\rangle_{S,Q}}}
\right)^{3/2}  {{1}\over {(1-\phi_{S,Q}^2(t))^{3/2}}}
\exp{\left[-{{3(X-\phi_{S,Q}(t)X_0)^2} \over {2\langle X^2\rangle_{S,Q}
(1-\phi_{S,Q}^2(t))}}\right]}
\end{equation}
specified by the second moment $\langle X^2\rangle_{S,Q}$ and the normalized correlation function 
$\phi(t)_{S,Q}$$=$${{\langle X(t)X(0)\rangle_{S,Q}}/{\langle X^2 \rangle_{S,Q}}}$. Calculations of 
$\langle X^2\rangle_{S,Q}$ and $\phi(t)_{S,Q}$ are given in the Appendix \cite{41,41a}. In the absence of 
force, we obtain:
\begin{equation}\label{4.5}
\langle X(t)X(0)\rangle_0 =12 k_B T \sum_{n=1}^{\infty}{{1}\over
{z_n}}\psi^2_n(L/2)  e^{-z_n t/\gamma},  \quad n=1,3,\ldots , 2q+1
\end{equation}
where $\psi_n(X)$ and $z_n$ are the eigenfunctions and eigenvalues of the modes of the operator that
describe the dynamics of ${\bf r}(s,t)$ (see Eq. (\ref{4.3})). To construct the propagator 
$G_{S,Q}(X,t;X_0)$ for ${\bf f}_{S,Q}$, Eq. (\ref{4.3}) is integrated with ${\bf f}_{S,Q}$ added to 
random force. We obtain: $\langle X^2 \rangle_{S,Q}$$=$$\langle X^2\rangle_0$$+$${\bf f}_{S,Q}^2
\sum_{n=1}^{\infty}\psi_n^2(L/2)/z_n^2 $, where  $n=1,3,\ldots , 2q+1$. We analyze the distributions of 
unfolding times $P(T,t)$ for the model sequence $S1$ (Figure 3) obtained using simulations, CTRW model 
for evolution of native interactions (Eqs. (\ref{3.2})-(\ref{3.5})) and Gaussian statistics of the 
protein coil (Eq. (\ref{4.2})). 

{\it Simulations of model $\beta$-sheet protein:} The usefulness of FCS is illustrated by computing and 
analyzing the distribution function $P(T;t)$ for a model polypeptide chain with $\beta$-sheet
architecture. Sequence $S1$, which is a variant of an off-lattice model introduced sometime ago \cite{20old}, 
is a coarse-grained model (CGM) of a polypeptide chain, in which each amino acid is substituted with a 
united atom of appropriate mass and diameter at the position of the $C_\alpha$-carbons \cite{20,21}. 
The $S1$ sequence is modeled as a chain of 46 connected beads of three types, hydrophobic $B$, 
hydrophilic $L$, and neutral $N$, with the contour length $L=46a$, where $a$$\approx$$3.8\AA$ is the 
distance between two consequtive $C_{\alpha}$- carbon atoms. The coordinate of $j$-th residue is given 
by the vector ${\bf x}_j$ with $j$$=$$1$, $2$, $\ldots$, $N$. 

The potential energy $U$ of a chain conformation is 
\begin{equation}
U = U_{bond} + U_{bend}+ U_{da} +  U_{nb},
\end{equation}
where $U_{bond}$, $U_{bend}$, $U_{da}$ are the energy terms, which
determine local protein structure, and $U_{nb}$ corresponds to non-local 
(non-bonded) interactions. The bond-length potential $U_{bond}$, that describes the
chain connectivity, is given by a harmonic function
\begin{equation}
U_{bond}=\frac{k_b}{2} \sum _{j=1}^{N-1} ( |{\bf X}_j -{\bf x}_{j+1}|-a)^2
\end{equation}
where $k_b$$=$$100$$\epsilon_h/a^2$ and $\epsilon_h$ ($\approx 1.25 kcal/mol$) 
is the energy unit roughly equal to the free energy of a hydropobic contact. 
The bending potential $U_{bend}$ is
\begin{equation}\label{V_BA}
U_{bend}=\sum_{j=1}^{N-2} \frac{k_{\theta }}{2}(\theta _j -  \theta _0)^{2},
\end{equation}
where $k_{\theta }$$=$$20$$\epsilon_{h}/rad^2$ and $\theta _{0}$$=$$105^{\circ}$. 
The dihedral angle potential $U_{da}$, which is largely responsible for maintaining 
protein-like secondary structure, is taken to be
\begin{equation}\label{V_DIH}
V_{da}=\sum_{i=1}^{N-3} [ A_{i}(1+\cos\phi_{i}) + B_{i}(1+\cos 3\phi_{i})],
\end{equation}
where the coefficients $A_i$ and $B_i$ are sequence dependent. Along the 
$\beta$-strands $trans$-states are preferred and $A$$=$$B$$=$$1.2$$\epsilon_h$. 
In the turn regions (i.e. in the vicinity of a cluster of $N$ residues) $A$$=$$0$, 
$B$$=$$0.2$$\epsilon_h$. The non-bonded 12-6 Lennard-Jones interaction $U_{nb}$ 
between hydrophobic residues is the sum of pairwise energies
\begin{equation}\label{Unb}
U_{nb} = \sum_{i<j+2} U_{ij}, 
\end{equation}
where $U_{ij}$ depend on the nature of the residues. The double summation in Eq. (\ref{Unb}) runs over all 
possible pairs excluding the nearest neighbor residues. The potential $U_{ij}^{BB}$ between a pair of 
hydrophobic residues $B$ is given by $U_{ij}^{BB}(r)$$=$$4$$\lambda$$\epsilon_{h}$
$\left[\biggl(\frac{a}{r}\biggr)^{12}-\biggl(\frac{a}{r}\biggr)^{6}\right]$,
where $\lambda$ is a random factor unique for each pair of $B$
residues \cite{21} and $r$$=$$|{\bf x}_i$$-$${\bf x}_j|$. For all other pairs
of residues $U_{ij}^{\alpha \beta}$ is repulsive \cite{21}. 

Although an off-lattice CGM drastically simplifies the polypeptide chain structure, it does 
retain important charateristics of proteins, such as chain connectivity and the heterogeneity 
of contact interactions. The local energy terms in S1 provide accurate representation of the 
protein topology. The native structure of $S1$ is a $\beta$-sheet protein that has a topology 
similar to the much studied immunoglobulin domains (Figure 3). When the model sequence is subject
to ${\bf f}_{S}$ or ${\bf f}_{Q}$, the total  energy is written as 
$U_{tot}$$=$$U$$-$${\bf f}_{\alpha}$${\bf X}$ ($\alpha$$=$$S$ or $Q$), where ${\bf X}$ 
is the protein end-to-end vector, and ${\bf f}_{S,Q}$$=$$(f_{S,Q},0,0)$ is applied 
along the ${\bf x}$-direction (Figure 1). 

The dynamics of the polypeptide chain is assumed to be given by the overdamped 
Langevin equation, which in the absence of ${\bf f}_S$ or ${\bf f}_Q$, is
\begin{equation}\label{2.14}
\eta {{d}\over {dt}}{\bf x}_j = - {{\partial U_{tot}}\over {\partial {\bf
x}_j}} + {\bf g}_j(t)
\end{equation}
where $\eta$ is the friction coefficient and ${\bf g}_j(t)$ is a Gaussian white noise,
with the statistics
\begin{equation}\label{2.15}
\langle {\bf g}_j(t)\rangle = 0, \quad  \langle {\bf g}_i(t){\bf
g}_j(t')\rangle = 6k_{B}T\eta \delta_{ij}\delta(t-t')
\end{equation}
Eqs. (\ref{2.14}) are integrated  with a step size $\delta t$$=$$0.02$$\tau_L$, 
where $\tau_L$$=$$(ma^2/\epsilon_h)^{1/2}$$=$$3 ps$ is the unit of time and 
$m\approx 3\times 10^{-22} g$ is a residue mass. In Eq. (\ref{2.14}) the value 
of $\eta$$=$$50 m/\tau_L$ corresponds roughly to water viscosity.

\section{Results}

{\it Simulations of unfolding and refolding of $S1$:} For the model sequence $S1$ we have previously 
shown that the equilibrium critical unfolding force is $f_C$$\approx $$22.6pN$ \cite{20} at the temperature
$T_s$$=$$0.692\epsilon_h/k_B$ below the folding transition temperature $T_F$$=$$0.7\epsilon_h/k_B$. 
At this temperature $70$$\%$ of native contacts are formed (see the phase diagram in Ref.~\cite{20}). 
To simulate the stretch-relax trajectories, the initially folded structures in the NBA were equilibrated  
for $60 ns$ at $T_s$. To probe forced unfolding of S1 at $T$$=$$T_s$, constant pulling force 
$f_{S}$$=$$40$$pN$ and $80pN$ was applied to both terminals of $S1$. For these values of $f_{S}$, $S1$ 
globally unfolds in $t$$=$$90 ps$ and $50 ps$, respectively. Cycles of stretching were interrupted by 
relaxation intervals during which the force is abruptly quenched to $f_{Q}$$=$$0$ for various duration $T$. 
Unfolding-refolding trajectories of $S1$ have been recorded as time series of $X$ and the number of native 
contacts $Q$.

In Figure 4 we present a single unfolding-refolding trajectory of $X$ and $Q$ of $S1$, generated by
stretch-relax cycles. Stretching cycles of constant force $f_{S}$$=$$80pN$ applied for $30 ns$ are 
interrupted by periods of quenched force relaxed over $90 ns$. A folding event is registered 
if it results in the formation of $92\%$ of the total number of native contacts $Q_F$$=$$106$, 
i.e. $Q$$\ge$$0.92Q_F$ for the first time. An unfolding time is defined as
the time of rupture of $92\%$ of all possible contacts for the first time. With this definition, the 
unfolded state end-to-end distance is $X$$\ge$$X_U$$\approx$$36$$a$. In Figure 4, folded (unfolded) states 
correspond to minimal (maximal) $X$ and maximal (minimal) $Q$. Inspection of Figure 4 shows that refolding 
events are essentially stochastic. Out of 36 relaxation periods only 9 attempts resulted in refolding of 
$S1$. Both $X$ and $Q$ show that refolding of $S1$ occurs though an initial collapse to a coiled state 
with the end-to-end distance $X_C/a$$\approx$$15$ ($Q$$\approx$$20$), followed by the establishment of 
additional native contacts ($Q$$\approx$$90$) stabilizing the folded state with $X_F/a$$\approx$$(1-2)$.

We generated about 1200 single unfolding-refolding trajectories and monitored the time-dependent
behavior of $X$ and $Q$. In the first set of simulations  we set $f_S$$=$$40pN$ and used several
values of $T$$=$$24$, $54$, $102$, $150$ and $240 ns$. In the second set, $f_S$$=$$80pN$, 
and $T$$=$$15$, $48$, $86$, $120$ and $180ns$. Each trajectory involves four stretching cycles 
separated by three relaxation intervals in which $f_Q$$=$$0$. Typical unfolding-refolding 
trajectories of $X$ and $Q$ for $f_S$$=$$40pN$, $f_Q$$=$$0$, and $T$$=$$102$, $150$ and 
$240 ns$ are displayed in Figures 5, 6 and 7, respectively. Due to finite duration of stretching 
cycles ($90 ns$), unfolding of $S1$ failed in few cases which were not included in the subsequent 
analysis of unfolding times. Only first stretching cycles in each trajectory are guaranteed to 
start from the NBA and for $T$$=$$102 ns$ (Figure 5) relatively few relaxation intervals 
result in refolding (with large $Q$). This implies that the distribution of unfolding 
times $P(T,t)$ obtained from these trajectories are dominated by contributions from the coiled 
states with the kinetics of formation of the native contacts playing only a minor role. Not unexpectedly, 
refolding events are more frequent when $T$ is increased to $150ns$ and $240 ns$. At $T$$=$$150ns$, 
$Q$ reaches higher values ($\approx$$65-75$) and the failure to refold is rare (Figure 6). 
This implies that as $T$ starts to exceed the (re)folding time $\tau_F$, the distribution of unfolding 
events, parametrized by $P(T,t)$, is characterized by diminishing contribution from the coiled states 
$\{ C \}$ and is increasingly dominated by the folded conformations in the NBA. Note that failed refolding 
events are observed even at $T$$=$$240 ns$ (Figure 7), which implies large heterogeneity in the 
duration of folding barrier crossing events. Figures 5-7 suggest that the folding time $\tau_F$ at 
the temperature $T_U$ is in the range $100$$-$$240ns$. Direct computations of the folding time $\tau_F$ 
from hundreds of folding trajectories starting with the fully stretched states gives 
${\bar \tau}_F$$\approx$$176ns$. The agreement between ${\bar \tau}_F$ and $\tau_F$ validates our 
stretch-release simulations. 

{\it Analysis of the distribution of unfolding times of $S1$:} The theoretical considerations in our 
formalism suggest that the $T$-dependent heterogeneous unfolding processes occur not only from the NBA but 
also from the intermediate coil $\{ C \}$ states. The $T$-dependent protein dynamics can be utilized to 
separately probe the coil dynamics of the polypeptide chain and the kinetics of formation/rupture of native 
contacts ($Q$). We now utilize unfolding-refolding trajectories of $S1$, simulated for short, intermediate 
and long $T$, to build the histograms of unfolding times $P(T,t)$. Using $P(T,t)$ we provide quantitative 
description of the polypeptide chain dynamics in the coil state and the kinetics of rupture and formation of 
native interactions by employing CTRW model for $Q$ and Gaussian statistics for $X$. 

We computed $P(T,t)$ using the distribution of unfolding times obtained for $f_S$$=$$80pN$, 
$T$$=$$15$, $48$ and $86ns$ (Figure 8), and $f_S$$=$$40pN$, $T$$=$$24$, $54$ and $102ns$ (Figure 9).
In both cases $f_Q$$=$$0$. We excluded unfolding times corresponding to the first 
stretch-quench cycle of each trajectory which were used to construct $P(t)$ for
the purposes of comparing $P(t)$ with $P(T,t)$ for long $T$. Single peaked $P(T,t)$ obtained for 
$T$$=$$15ns$ (Figure 8) and $T$$=$$24ns$ (Figure 9), represent contributions to $S1$ unfolding 
from coil manifold $\{ C\}$ alone. When $T$ is increased to $48ns$ (Figure 8) and $54ns$ 
(Figure 9), position of the peak shifts to longer times, i.e. from $t$$\approx$$2.5ns$ 
to $t$$\approx$$5ns$ (Figure 8) and from $t$$\approx$$6ns$ to $t$$\approx$$10ns$ (Figure 9). 
Furthermore, $P(T,t)$ develops a shoulder at $t$$\approx$$10ns$ and $t$$\approx$$25ns$, observed for 
$T$$=$$86ns$ (Figure 8) and $T$$=$$102ns$ (Figure 9), which indicates a growing 
(with $T$) contribution to unfolding from relaxation trajectories that reach the NBA. At longer 
$T$$=$$150ns$, when most relaxation periods result in refolding of $S1$, contribution from 
coiled states diminishes and at $T$$=$$240ns$ $P(T,t)$ is identical to the standard 
distribution $P(t)$ constructed from unfolding times of the first stretch-quench cycle 
of each trajectory. This implies that for $f_Q$$=$$0$, $\tau_F$$\approx$$240ns$ and that 
$P(T,t)$$\to$$P(t)$ for $T$$>$$240ns$. The distribution $P(T,t)$$=$$P(t)$ constructed from 
unfolding times separated by $T$$=$$300ns$ is presented in Figures 8 and 9 (top left panel).

We use the CTRW formalism to analyze the histograms of unfolding times $P(T,t)$ from which the parameters
that characterize the energy landscape of $S1$ can be mapped. We describe the kinetics of rupture and formation 
of native contacts by the waiting time distributions $\Psi_r$, $\Psi_f$, 
\begin{equation}\label{5.1}
\Psi_r (t)= N_r t^{v_r-1}e^{-k_r t}, \qquad \Psi_f(t)= N_f t^{v_f-1}e^{-k_f t}
\end{equation}
where $k_r$ (dependent on $f_S$) and $k_f$ (dependent on $f_Q$) are the rates of rupture and formation 
of native interactions, respectively, $N_{r,f}$$=$$k_{r,f}/\Gamma(v_{r,f})$ are normalization constants 
($\Gamma(x)$ is Gamma function) and $v_{r,f}$$\ge$$1$ are phenomenological parameters quantifying the
deviations of the kinetics from a Poissonian process. For instance, $v_{r,f}$$=$$1$ implies Poissonian
process and corresponds to standard chemical kinetics with constant rate $k_{r,f}$. We assume that both the 
folded and the unfolded states are sharply distributed around the mean native and unfolded end-to-end 
distance $\langle X_F\rangle$ and  $\langle X_U\rangle$, respectively (Figure 2),
\begin{equation}\label{5.2}
P_{eq}(X_F)=\delta (X-\langle X_F\rangle), \quad \text{and} \quad
P(X_U)=\delta (X-\langle X_U\rangle)
\end{equation}
where $\langle X_U\rangle/a$$=$$36$ residues corresponds to the definition of unfolded state. For $S1$ 
the contour length $L/a$$=$$46$. Thus, $S1$ is unfolded if $X/a$ exceeds $\langle X_U \rangle $ which 
implies $\delta /a$$=$$10$ residues (see Figure 2 and the lower limit of integration in Eq. (\ref{2.1})). 
We describe the distribution of states $\{ C\}$ before the transition to the NBA by a Gaussian,
\begin{equation}\label{5.3}
P_C(X)=e^{-(X-\langle X_C\rangle)^2/2\Delta X_C^2}
\end{equation}
with the width $\Delta X_C$, centered around the average distance $\langle X_C\rangle$.

We performed numerical fits of the histograms presented in Figures 8 and 9 using  Eqs. (\ref{2.1}), 
(\ref{2.2}), (\ref{2.4}) and (\ref{2.5}). By fitting the theoretical curves to $P(T,t)$ constructed from 
short $T$$=$$15ns$ and $T$$=$$48ns$ simulations (Figure 8) and $T$$=$$24ns$ and $T$$=$$54ns$ (Figure 9), 
we first studied the dynamics of $X$ to estimate the dynamical timescale $\tau_d$, i.e. the longest 
relaxation time corresponding to the smallest eigenvalue $z_n$ (Eq. (\ref{4.5})), and persistence 
length $l_p$ of $S1$ in the coil states $\{ C \}$. By using the values of $\tau_d$ and $l_p$, we used 
our theory to describe $P(T,t)$ constructed from long $T$$=$$300ns$ simulations. This analysis allows 
us to estimate the parameters characterizing the rupture of native contacts $k_r$, $v_r$, 
$\langle X_F\rangle$ and $\Delta X_F$. Finally, the parameters $k_f$, $v_f$,  $\langle X_C \rangle$ and 
$\Delta X_C$, characterizing formation of native contacts were estimated using $\tau_d$, $l_p$, $k_r$, 
$v_r$, $\langle X_F\rangle$ and $\Delta X_F$, and fitting Eqs. (\ref{2.4}) and (\ref{2.5}) to $P(T,t)$ 
for intermediate $T$$=$$86ns$ (Figure 8) and $T$$=$$102ns$ (Figure 9). 

{\it Extracting the energy landscape parameters of $S1$:} There are a number of parameters that characterize
the energy landscape and the dynamics of the major components in the NBA$\to$U transition. The numerical 
values of the model parameters are summarized in the Table. The values of $v_r$$=$$6.9$ for  $f_S$$=$$40pN$ 
and $v_r$$=$$5.1$ for $f_S$$=$$80pN$ indicate that rupture of native contacts is highly cooperative especially 
at the lower $f_S$$=$$40pN$. This agrees with the previous findings on kinetics of forced unfolding of $S1$ 
\cite{20} which were based solely on unfolding $S1$ by applying a constant force. In contrast, the formation 
of native contacts is characterized by $v_f$$\approx$$1$ implying almost Poissonian distribution for the 
kinetics of formation of native contacts. The structural characteristics of the coil states are obtained 
using the relaxation of the polypeptide chain upon force-quench from stretched states. The value of the 
persistence length $l_p$, which should be independent of ${\bf f}_Q$ provided ${\bf f}_Q$$/$${\bf f}_C$$\ll$$1$, 
is found to be about $4.8$$\AA$ (Table). This value is in accord with the results of the recent experimental 
measurements based on kinetics of loop formation in denatured states of proteins \cite{41b}. 

Upon rupture of native contacts, the chain extends by $\Delta X_F/a$$=$$6.4$ (for $f_S$$=$$40pN$) and 
$\Delta X_F/a$$=$$6.7$ (for $f_S$$=$$80pN$). This distance separates the basins of folded states with 
$\langle X_F\rangle /a$$=$$4.5$ at $f_S$$=$$40pN$ and $\langle X_F\rangle /a$$=$$4.6$ at $f_S$$=$$80pN$ 
from high free energy states when the polypeptide chain is stretched in the direction of ${\bf f}_S$ 
(Figure 2(a)). Because these high free energy states are never populated we expect that forced-unfolding
of $S1$ must occur in an apparent two-step manner when $T$$\to$$\infty$. Explicit simulations of $S1$
unfolding at constant ${\bf f}_S$ ($\approx$$69pN$) shows that mechanical unfolding occurs in a single step
(see Figure 2 in Ref.~\cite{20}). 

From the refolding free energy profile upon force-quench (see Figure 2(b)) we infer that the initial
stretched conformation must collapse to an ensemble of compact structures $\{ C \}$. From the analysis
of $P(T;t)$ using the CTRW formalism we find that the average end-to-end distance $\langle X_C\rangle$ for 
the manifold $\{ C \}$ is close to $\langle X_F \rangle$ (see the Table) which suggests that the ensemble of 
the $\{ C \}$$\to$ NBA transition states is close to the native state. There is a broad distribution of 
coiled states $\{ C \}$ which is manifested in the large width $\Delta X_C/a$$=$$2.2$. Due to the broad 
conformational distribution, there is substantial heterogeneity in the refolding pathways. This feature 
is reflected in the long tails in $P(T,t)$ (see Figures 8 and 9). As a result, we expect the kinetic 
transition to be sharp. The estimated timescale ($\sim$$1/k_f$) for forming native contacts for $S1$ is 
shorter than the coil dynamical timescale $\tau_d$ (for the values of $f_S$ used in the simulations). 
This indicates that the dynamical collapse of $S1$ from the stretched state $X_U$$\approx$$L$ and equilibration 
in the coiled manifold $\{ C \}$ constitites a significant fraction of the total folding time 
($\approx$$\tau_d$$+$$k_f^{-1}$). From the analysis of folding of $S1$ ($P(T;t)$ at intermediate $T$) we
also infer that the transition state ensemble for $\{ C \}$$\to$$N$ must be narrow.

From the rates of rupture of native contacts $k_r$ at the two ${f}_S$ values and assuming Bell model 
for the dependence of $k_r$ on $f_S$,
\begin{equation}\label{5.4}
k_r(f_S) = k_r^0 e^{\sigma f_S/k_B T}
\end{equation}
we estimated the force-free rupture rate $k_r^0$ and the critical extension $\sigma$ at which folded 
states of $S1$ become unstable. We found that $k_r^0$$=$$8$$\times$$10^{-4}ns^{-1}$ is negligible 
compared to the rate of formation of native contacts, $k_f$$=$$0.25ns^{-1}$. The location of the transition
state of unfolding $X$$=$$\langle X_F\rangle$$+$$\sigma$ is characterized by 
$\sigma$$=$$1.5$$a$$\approx$$0.03$$L$. The value of $\sigma$ is short compared to $\Delta X_C$ which is 
a measure of the width of the $\{ C \}$ manifold. Small $\sigma$ implies that the major barrier to unfolding
is close to the native conformation. A similar values of $\sigma$ was obtained in the previous study
of $S1$ by using an entirely different approach \cite{20}. These findings are consistent with AFM 
experiments \cite{Rief97Science} and computer simulations \cite{KlimThirum99PNAS} which show that 
native structures of proteins appear to be ``brittle'' upon application of mechanical force. 

The parameter $\tau_d$ is an approximate estimate of the collapse time, $\tau_c$, from the stretched 
to the coiled state. Using direct simulations of the decay of the radius of gyration, $R_g$, starting 
from a rod-like conformation, we obtained $\tau_c$$\approx$$80ns$ (see Supplementary Information in
\cite{44}). The value of $\tau_d$ ($\approx$$20ns$) is in reasonable agreement with the estimate
of $\tau_c$. This exercise shows that reliable estimates of timescales of conformational dynamics, which 
are difficult to obtain, can be made using FCS. To ascertain the extent to which the estimate of $K_U$ agrees 
with independent calculations, we obtained the $K_U$ by applying a constant force to unfold $S1$. The value 
of $K_U$, obtained by averaging over $200$ trajectories, is about $90ns$ at $f_S$$=$$40pN$ which is in 
rough accord with $K_U$$\approx$$\tau_d$$+$$k_r^{-1}$$\approx$$70ns$. This further validates the efficacy 
of FCS in obtaining the energy landscape of proteins. We also estimated $K_U^0$ from the value of $K_U$ 
obtained by direct simulation and the Bell model. The ${\bf f}_S$-dependent unfolding rate 
$K_U$$\approx$$\tau_d$$+$$k_r^{-1}$ increases with ${\bf f}_S$ in accord with Eq. (\ref{5.4}). The 
prefactor ($K_U^0$) is about ten fold smaller than $k_r^0$. The difference may be either due to the 
failure of the assumption that $k_r^0$$=$$K_U^0$ or the breakdown of the Bell model \cite{45}.

\section{Discussion}

In this Section we summarize the main steps for practical implementation of the proposed Force Correlation 
Spectroscopy (FCS) to probe the energy landscape of proteins using forced unfolding of proteins. 

{\it Step 1. Evaluating the (re)folding timescale $\tau_F$:} In the first phase of the FCS experiments,
one needs to collect a series of histograms $P(T_n,t)$, $n$$=$$1,2,\ldots$, $N$ of unfolding times for 
increasing relaxation time $T_1$$<$$T_2$$<$$\ldots$$<$$T_N$ by repeated stretch-release experiments. 
This can be done by discarding the first unfolding time $t_1$ in the sequence of recorded unfolding 
times $\{ t_1, t_2, \ldots, t_M \}$ for each $T_n$ to guarantee that all the unfolding events are generated 
from the stretched states with the distribution $P(X_U)$ (see Eq. (\ref{2.1})). This is a crucial element 
of the FCS methodology since it enables us to perform the averaging over the final (stretched) states.
It is easier to resolve experimentally the end-to-end distance $X$$\approx$$L$, rather than the initial 
(folded) states in which a number of conformations belong to the NBA. The histograms are compared with 
$P(T^*,t)$ obtained for sufficiently long $T^*$$\gg$$\tau_F$. To ensure that $T^*$ exceeds $\tau_F$, 
$T^*$ can be as long as few tens of minutes. The time at which $D(T_n)$, given by Eq. (\ref{2.3}), 
is equal to zero can then be used to estimate $\tau_F$. Notice that our estimate of $\tau_F$ 
{\it does not} hinge on whether $P(T\to\infty;t)$$\equiv$$P(t)$ is Poissonian or not! Clearly, the 
choice of $T^*$ depends on the protein under the study, and prior knowledge or bulk measurements of 
unfolding times observed under the influence of temperature jump or denaturing agents can serve as a 
guide to estimate the order of magnitude of $T^*$. 

{\it Step 2. Resolving the dynamics of the polypeptide chain:} To this end we have determined the
ensemble average (re)folding time, $\tau_F$. In the second phase of the FCS, we perform statistical analysis 
of the distribution of unfolding times collected at $T$$\ll$$\tau_F$, i.e. $P(T\ll \tau_F;t)$ 
(regime I in Section II). This allows us to probe the dynamic properties of the polypeptide chain, 
such as the protein persistence length $l_p$ and the protein dynamical timescale $\tau_d$ (see the Table). 
Indeed, by assuming a reasonable model for the conditional probability, $G(X',t;X)$, of the protein 
end-to-end distance and the distribution of the stretched states, $P(X_U)$,  $l_p$ and $\tau_d$ can be 
determined from the fit (either analytically or numerically) of the unfolding time distribution, 
$P(T\ll\tau_F;t)$, given by Eq. (\ref{2.1}), to the histogram of unfolding times collected for 
$T$$\ll$$\tau_F$. To illustrate the utility of the FCS, in the present work we assumed a Gaussian profile 
for $G_{S,Q}(X',t;X)$ (see Eq. (\ref{4.2})) and the worm-like-chain model for the polypeptide chain. The 
general formulae (\ref{2.1}) allows for the use of more sophisticated models of $X$, should it become 
necessary. Recent single molecule FRET experiments on proteins \cite{46,47}, dsDNA, ssDNA, and RNA \cite{48} 
show, surprisingly, that the characteristics of unfolded states obey worm-like chain models. Moreover,
all the data in forced unfolding of proteins have been analyzed using WLC models. Thus, the analysis
of FCS data using WLC dynamics for unfolded polypeptide chains to a large extent is justified. $G_S(X',t;X)$ 
and $G_Q(X',t;X)$ can be ``measured'' in the current AFM and LOT experiments by computing the frequency 
of occurence of the event $X$ after the forced stretch (${\bf f}$$=$${\bf f}_S$) or force quench 
(${\bf f}$$=$${\bf f}_Q$) from the well-controlled partially stretched state $X$ or the fully stretched 
state $X$$\approx$$L$ of the chain, respectively, over time $t$ ($\ll$$\tau_F$). 

{\it Step 3. Probing the kinetics of rupture of the protein native contacts:} Having resolved the dynamics
of the protein in extension-time regime, where the number of native interactions that stabilize the native 
state is small, we can resolve the kinetics of forced rupture of native interactions stabilizing the NBA 
(regime II). In the third part of the FCS we analyze the distribution of unfolding times for 
$T$$\gg$$\tau_F$, given by Eq. (\ref{2.2}). We use the knowledge about the propagator $G_S(X',t';X,t)$, 
appearing in the rhs of Eq. (\ref{2.2}), obtained in {\it Step 2} to perform analytical or numerical fit 
of the distribution $P(T\gg \tau_F;t)$ to the histogram of unfolding times collected for $T$$\gg$$\tau_F$. 
The new information, gathered in {\it Step 3}, sheds the light on the kinetics of native interactions 
stabilizing the NBA, which is contained in the probability $P_F(t;{\bf f}_{S},X_F)$ (see Eq. (\ref{2.2})). 
For convenience, we used the continuous time random walk (CTRW) model for $P_F(t;{\bf f}_{S},X_F)$, 
which is summarized in Eqs. (\ref{3.3})-(\ref{3.5}), and the assumption of separability, given by 
Eqs. (\ref{3.2}). CTRW reduces to the Poissonian kinetics with the rate constants when the waiting time 
distribution function for the rupture of native contacts, $\Psi_r(t)$, is an exponential function of $t$. 
The CTRW probes the possible deviations of the kinetics of $P_F(t;{\bf f}_{S},X_F)$ from the Poisson process 
and allows to test different functional forms for $\Psi_r(t)$. In the simplest implementation
of CTRW utilized in the present work, $\Psi_r(t)$ is assumed to be an algebraic function of $t$, given by Eqs.
(\ref{5.1}), which allows us to estimate the rate of rupture of native interactions, $k_r$, and parameter
$v_r$ quantifying the deviations of the rupture kinetics from a Poissonian process. Furthermore, by repeating
{\it Step 3} for different values of the stretching force, $f_S$, and assuming the Bell model for $k_r(f_S)$,
given by Eq. (\ref{5.4}), we can also estimate the force-free rupture rate, $k_r^0$, and the critical extension
$\sigma$, which quantifies the distance from the NBA to the transition state along the direction of $f_S$. 
We also obtain the average end-to-end distance in the folded state, $\langle X_F\rangle$ from the distribution 
of the native states $P_{eq}(X_F)$. 

{\it Step 4. Resolving the kinetics of formation of native interactions:} In the final step the 
distributions $P(T\ll \tau_F;t)$ and $P(T\gg \tau_F;t)$, analyzed in {\it Steps 2} and {\it 3} respectively, 
are used to form a linear superposition (Eq. (\ref{2.4}), regime III). The $T$-dependent weights 
are given by the probabilities $\rho_C(T)$ and $\rho_F(T)$$=$$1$$-$$\rho_C(T)$, respectively. This 
superposition is used to fit the histogram of unfolding times, $P(T\sim \tau_F;t)$, collected for 
$T$$\sim$$\tau_F$. The estimated probability $\rho_F(T)$ should then be matched with the propbability 
obtained by performing double integration in Eq. (\ref{2.5}). This allows us to probe the kinetics of 
formation of native contacts, $P_C(T;X,{\bf f}_Q)$, for the known propagator $G_Q(X',T;X)$ analyzed in 
{\it Step 2}. As in the case of $P_F(t;{\bf f}_{S},X_F)$, we assumed separability condition for 
$P_C(t;{\bf f}_{Q},X_C)$ (Eqs. (\ref{3.2})) and CTRW for the kinetics of formation of native contacts 
contained in $P_f(t;{\bf f}_Q)$ (see Eqs. (\ref{3.3})-(\ref{3.5})). A simple algebraic form for the 
waiting time distribution function, $\Psi_f(t)$, given by Eq. (\ref{5.1}), allows us to estimate the 
force-free rate of formation of native interactions, $k_f(f_Q=0)$$=$$k_f^0$. Moreover, the heterogeneity 
of the protein folding pathyways can be assesses by analyzing the width, $\Delta X_C$, of the distribution 
of coiled protein states, $P_C(X)$, centered around the average end-to-end distance, $\langle X_C\rangle$ 
(see Eq. (\ref{5.3})). Similar to the analysis of rupture kinetics, {\it Step 4} could be repeated for the 
two values of the quenched force, $f_Q$, to yield the force-free rate of formation of native contacts, 
stabilizing the native fold, and the distance between $\langle X_C\rangle$ and the transition state for 
the formation of native contacts. For the purposes of illustration, in the present work we used $f_Q$$=$$0$.

At the minimum FCS can be used to obtain model-independent estimate of $\tau_F$. By assuming a WLC
description for coiled states, which is justified in light of a number of FRET and forced unfolding 
experiments, estimates of collapse times and their distribution as well as persistence length can be
obtained. If CTRW model is assumed then estimates of timescale for rupture and formation of native
contacts can be made. The utility of FCS for $S1$ illustrates the efficacy of the theory. The potential
of obtaining hitherto unavailable information makes FCS extremely useful.

\section{Conclusions}

In this paper we have developed a theory to describe the role of internal relaxation of polypeptide chains 
in the dynamics of single molecule force-induced unfolding and force-quench refolding. To probe the effect 
of dynamics of the chain in the compact manifold of states, that are populated in the pathways to the NBA 
starting from the stretched conformations, we propose using a series of stretch-release cycles. In this new 
class of single molecule experiments, referred to as force correlation spectroscopy (FCS), the duration of 
release times ($T$) is varied. FCS is equivalent to conventional mechanical unfolding experiments in the limit
$T$$\to$$\infty$. By applying our theory to a model $\beta$-sheet protein we have shown that the parameters 
that characterize the energy landscape of proteins can be obtained using the joint distribution function of
unfolding times $P(T;t)$. 

The experimentally controllable parameters are ${\bf f}_{S}$, ${\bf f}_{Q}$, and $T$. In our illustrative 
example, we used values of ${\bf f}_{S}$ that are approximately $(2-4)$ times greater than the equilibrium 
unfolding force. We set ${\bf f}_{Q}$$=$$0$ which is difficult to realize in experiments. From the schematic 
energy landscape in Figure 1 it is clear that the profiles corresponding to the positions of the manifold 
$\{ C\}$, the dynamics of $\{ C\}$, and the transition state location and barrier hight depend on ${\bf f}_{Q}$. 
The simple application, used here for proof of principle purposes only, already illustrates the power of FCS. 
To obtain the energy landscape of $S1$ by using FCS that covers a broader range of ${\bf f}_{S}$ and 
${\bf f}_{Q}$, a complete characterization of the landscape can be made. The experiments that we propose based 
on the new theoretical development can be readily performed using presently available technology. Indeed, 
the pioneering experimental setup used by Fernandez and Li \cite{13a} that have utilized force to initiate 
refolding can be readily adopted to perform single molecule FCS.

It is known that even for proteins that fold in an apparent two-state manner the energy landscape is
rough \cite{12b}. The scale of roughness $\Delta E$ can be measured in conventional AFM experiments by 
varying temperature. The extent to which the internal dynamics of proteins is affected by $\Delta E$, 
whose value is between $(2-5)$$k_B T$ \cite{50,51}, on the force-quenched refolding is hard to predict. 
These subtle effects of the energy landscape can be resolved (in principle) using FCS in which temperature 
is also varied. 

\acknowledgments{This work was supported in part by a grant from the National Science Foundation 
through grant number NSF CHE-05-01456.}

\appendix

\section{Calculation of $\langle X(t)X(0)\rangle $ }

In this Appendix we outline the calculation of $\langle X(t)X(0)\rangle $ and $\langle X^2\rangle $ 
for the force-free propagator $G_0(X,t;X_0)$. By using Eq. (\ref{4.1}) (without the last term) and 
applying the least action principle to WLC Lagrangian 
$L$$=$$m/2$$\int_{-L/2}^{L/2}ds$$(\partial{\bf r}/\partial t)^2$$-$$H$, we obtain: 
$m$${{\partial^2}\over {\partial t^2}}$${\bf r}(s,t)$$+$$\epsilon$${{\partial^4} \over {\partial s^4}}$
${\bf r}(s,t)$$-$$2$$\nu$${{\partial^2}\over {\partial s^2}}$ ${\bf r}(s,t)$$=$$0$, where $m$ is the 
protein segment mass and $\epsilon$$=$${{3l_p k_B T}/ {4}}$, $\nu$$=$${{3 k_B T}/{2l_p}}$. Dynamics of 
the media is taken into account by including a stochastic force $f(s,t)$ with the white noise statistics, 
$\langle{f}_{\alpha} (s,t) \rangle$$=$$0$, $\langle{f}_{\alpha}(s,t){f}_{\beta}(s',t')\rangle$$=$
$2$$\gamma$$k_BT$$\delta_{\alpha \beta}$$\delta(s-s')$$\delta(t-t')$, where $\alpha$$=$$x$, $y$, $z$, 
and $\gamma $ is the friction coefficient per unit coil length. In the overdamped limit, the equation
of motion for ${\bf r}(s,t)$ is \cite{41,41a}
\begin{equation}\label{4.3}
\gamma {{\partial}\over {\partial t}}{\bf r}(s,t) + \epsilon
{{\partial^4} \over {\partial s^4}}{\bf r}(s,t) -2\nu
{{\partial^2}\over {\partial s^2}}{\bf r}(s,t) = {\bf f}(s,t)
\end{equation}
with the boundary conditions,
\begin{equation}\label{4.4}
\left[ 2\nu {{\partial}\over {\partial s}}{\bf r}(s,t)-\epsilon
{{\partial^3} \over {\partial s^3}}{\bf r}(s,t)  \right]_{\pm L/2} = 0, 
\qquad \left[ 2\nu_0 {{\partial}\over {\partial s}}{\bf
r}(s,t)+\epsilon {{\partial^2} \over {\partial s^2}}{\bf r}(s,t)
\right]_{\pm L/2} = 0
\end{equation}
where $\nu_0$$=$$3k_B T/4$. We solve Eq. (\ref{4.3}) by expanding ${\bf r}(s,t)$ and ${\bf f}(s,t)$ in a 
complete set of orthonormal eigenfunstions $\{\psi_n(s)\}$, i.e.
\begin{equation}\label{A.1}
{\bf r}(s,t) =  \sum_{n=0}^{\infty} {\bf \xi}_n (t) \psi_n(s)\quad
\text{and} \quad {\bf f}(s,t) =  \sum_{n=0}^{\infty} {\bf f}_n (t)
\psi_n(s)
\end{equation}
Substituting Eqs. (\ref{A.1}) into Eq. (\ref{4.3}) and separating variables we obtain:
\begin{equation}\label{A.2}
\epsilon {{d^4}\over {ds^4}}\psi_n(s)-2\nu {{d^2}\over
{ds^2}}\psi_n(s) =  z_n \psi_n(s) \quad \text{and} \quad \gamma
{{d}\over {dt}}{\bf \xi}_n(t) + z_n {\bf \xi}_n (t) = {\bf f}_n(t)
\end{equation}
where $z_n$ is the $n$-th eigenvalue. The second Eq. (\ref{A.2}) for ${\bf \xi}(t)$  is solved by
\begin{equation}\label{A.3}
{\bf \xi}_n(t)={{1}\over {\gamma}}\int_{-\infty}^{t}dt' {\bf f}_n(t')
\exp{\left[-{{(t-t')z_n}\over {\gamma}}\right]}
\end{equation}
and the eigenfunctions $\psi_n(s)$ are
\begin{eqnarray}\label{A.4}
\psi_0 & = & \sqrt{{{1}/ {L}}}\\\nonumber \psi_n(s) & = &
\sqrt{{{c_n}/ {L}}}\left(  {{\alpha_n}\over {\cos{[\alpha_n L/2]}}}
\sin{[\alpha_n s]}  + {{\beta_n}\over {\cosh[\beta_n L/2]}}
\sinh{[\beta_n s]}\right), n=1,3,\ldots , 2q+1  \\\nonumber \psi_n(s)
& = & \sqrt{{{c_n}/ {L}}}\left( -{{\alpha_n}\over {\sin{[\alpha_n
L/2]}}} \cos{[\alpha_n s]} +{{\beta_n}\over {\sinh{[\beta_n
L/2]}}}\cosh{[\beta_n s]}\right), n=2,4,\ldots , 2q
\end{eqnarray}
where $c_n$'s are the normalization constants; $\alpha_n$ and $\beta_n$ are determined from 
Eqs. (\ref{4.4}),
\begin{eqnarray}\label{A.5}
& \alpha_n & \sin{[\alpha_n L/2]}\cosh{[\beta_n L/2]}-\beta_n^3
\cos{[\alpha_n L/2]} \sinh{[\beta_n L/2]} \\\nonumber & - &  {{1}\over
{l_p}}(\alpha_n^2+\beta_n^2)\cos{[\alpha_n L/2]}\cosh{[\beta_n
L/2]}=0,  \quad n=1,3, \ldots , 2q+1  \\\nonumber & \alpha_n &
\cos{[\alpha_n L/2]}\sinh{[\beta_n L/2]}+\beta_n^3 \sin{[\alpha_n
L/2]} \cosh{[\beta_n L/2]} \\\nonumber & + &  {{1}\over
{l_p}}(\alpha_n^2+\beta_n^2)\sin{[\alpha_n L/2]}\sinh{[\beta_n
L/2]}=0,  \quad n=2,4, \ldots , 2q
\end{eqnarray}
The parameters $\alpha_n$ and $\beta_n$ are related as $\beta_n^2-\alpha_n^2 = {{1}\over {l_p^2}}$. The
eigenvalues  $z_n$ are given by $z_n$$=$$\epsilon$$\alpha_n^4$$+$$2$$\nu$$\alpha^2$. Using
Eqs. (\ref{A.1}) and (\ref{A.3}), we obtain:
$\langle {\bf r}(s,t){\bf r}(s',t)\rangle=3k_B T\sum_{n=0}^{\infty} {{1}\over {z_n}}
\psi_n(s)\psi_n(s')e^{-z_nt/\gamma}$. Then,  
$\langle X(t)X(0)\rangle$$=$ $\langle{\bf r}({{L}\over
{2}},t){\bf r}({{L}\over {2}},0)\rangle$$+$ $\langle{\bf r}(-{{L}\over
{2}},t){\bf r}(-{{L}\over {2}},0)\rangle$$-$ $\langle{\bf r}({{L}\over
{2}},t){\bf r}(-{{L}\over {2}},0)\rangle$$-$ $\langle{\bf
r}(-{{L}\over {2}},t){\bf r}({{L}\over {2}},0)\rangle$,  which yields Eq. (\ref{4.5}).

\clearpage

\begin{table}
\caption{Energy landscape parameters for $S1$ extracted from FCS:}
\bigskip
\bigskip
\begin{minipage}{\textwidth}
\begin{tabular}{|c|c|c|c|c|c|c|c|c|c|c|}
\hline
$f_S$, $pN$\footnote{$f_S$ is the magnitude of the stretching
force} &$l_p/a$\footnote{$l_p$ is the persistence length of $S1$ in the coiled state (Eq. (\ref{4.1})) 
measured in units of $a$ ($\approx$$\AA$)} &$\tau_d$, $ns$\footnote{$\tau_d$ is the $f_Q$-dependent longest 
relaxation time in the coil state (Eq. (\ref{4.5}))} &$k_r$, $1/ns$\footnote{$k_r$ ($k_f$) is the rate 
of rupture (formation) of native interactions (Eq. (\ref{5.1})) and is a function of $f_S$ ($f_Q$)} 
& $\nu_r$\footnote{$\nu_{r}$ ($\nu_f$) quantifies deviations of the native contacts rupture (formation) 
kinetics from the Poisson process} 
&$\langle X_F\rangle /a$\footnote{$\langle X_F \rangle$ ($\langle X_C \rangle$) 
is the average end-to-end distance of $S1$ in the NBA (manifold $\{ C \}$) (Figure 2(b), 
Eqs. (\ref{5.2})-(\ref{5.3}))} &$\Delta X_F /a$\footnote{$\Delta X_F$ is the 
extension of the chain prior to rupture of all native contacts (Figure 2(a) and Eq. (\ref{2.2}))} 
&$k_f$, $1/ns$\footnote{$k_r$ ($k_f$) is the rate
of rupture (formation) of native interactions (Eq. (\ref{5.1})) and is a function of $f_S$ ($f_S$)} 
&$\nu_f$\footnote{$\nu_{r}$ ($\nu_f$) quantifies deviations of the native contacts rupture (formation) 
kinetics from the Poisson process} 
&$\langle X_C\rangle /a$\footnote{$\langle X_F \rangle$ ($\langle X_C \rangle$) 
is the average end-to-end distance of $S1$ in the NBA (manifold $\{ C \}$) 
(Figure 2(b), Eqs. (\ref{5.2})-(\ref{5.3}))} &$\Delta X_C /a$\footnote{$\Delta X_C$ is the width of 
the distribution of coiled states of $S1$ (Eq. (\ref{5.4})), a measure of the refolding heterogeneity} \\
\hline
40 &1.2 &19.6 &0.02 &6.9 &4.5 &6.4 &0.26 &1.1 &4.8 &2.2 \\
\hline
80 &1.1 &15.2 &0.11 &5.1 &4.6 &6.7 &0.25 &1.1 &4.7 &2.2 \\
\hline
\end{tabular}
\label{UCTable.ps}
\end{minipage}
\end{table}

\clearpage

\section*{\bf FIGURE CAPTIONS}

{\bf Figure 1.} $a$: A typical AFM setup: constant force ${\bf f}$$=$${\bf f}_{S}$$=$$f_S$${\bf x}$ 
is applied through the cantilever tip linker in the direction ${\bf x}$ parallel to the protein 
end-to-end vector ${\bf X}$. Stretching cycles are interrupted by relaxation intervals $T$ during which 
the force is quenched, ${\bf f}$$=$${\bf f}_{Q}$$=$$f_Q$${\bf x}$ ($f_S$$>$$f_Q$). $b$: A single trajectory 
of forced unfolding times $t_1$,  $t_2$, $t_3$, $\ldots$, separated by fixed relaxation time $T$, during 
which the unfolded protein can either collapse into the manifold of colied states $\{ C \}$ if $T$ is
short or reach the native basin of attraction (NBA) if $T$ is long.

\bigskip

{\bf Figure 2.} Schematic of the free energy profile of a protein (red) upon stretching at constant force 
${\bf f}_S$ and force-quench ${\bf f}_Q$. (a): The projections of energy landscape (blue) is in the direction 
of ${\bf X}$ which is a suitable reaction coordinate for unfolding induced by force ${\bf f}_S$. The average 
end-to-end distance in the native basin of attraction is $\langle X_F \rangle$. Upon application of 
${\bf f}_{S}$, rupture of contacts that stabilize the folded state $F$ results in the formation of an ensemble 
of high energy extended (by $\Delta X_F$) conformations $\{ I \}$. Subsequently, transitions to globally 
unfolded state $U$ (with $L-\delta$$\le$$X$$\le$$L$) occurs. (b): Free energy profile for force-quench 
refolding which occurs in the order $U$$\to$$\{ C \}$$\to$$F$. Refolding is initiated by quenching the force 
${\bf f}_S$$\to$${\bf f}_Q$$<$${\bf f}_C$, where ${\bf f}_C$ is the equilibrium critical force needed to unfold
the native protein. The initial event in the process is the formation of an ensemble of compact structures.
The mean end-to-end distance of $\{ C \}$ is $\langle X_C\rangle$ and the width is $\Delta X_C$ which is a 
measure of heterogeneity of the refolding pathways. These states may or may not end up in the native basin 
of attraction (NBA) depending on the duration of $T$. We have used ${\bf X}$ as a reaction coordinate during 
force-quench for purposes of illustration only.

\bigskip
{\bf Figure 3.} Native structure of the model protein $S1$. The model polypeptide chain has a $\beta$-sheet 
architecture of the native state. The $\beta$-strands of the model chain are formed by native contacts between 
hydrophobic residues (given by blue balls). The hydrophilic residues are shown by red balls and the residues 
forming the turn regions are given in grey.
 
\bigskip

{\bf Figure 4.} A single unfolding-refolding trajectory of the end-to-end distance $X/a$ (black) and the total 
number of native contacts $Q$ (red) as a function of time $t$ for $S1$. The trajectory is obtained by repeated 
application of stretch-quench cycles with stretching force $f_{S}$$=$$80 pN$ and quenched force $f_Q$$=$$0$. 
The duration of streching cycle and relaxation period is $30ns$ and $90ns$, respectively. The first five 
unfolding events corresponding to large $X/a$ and small $Q$ are marked explicitely by numbers $1$, $2$, $3$, 
$4$ and $5$. Force stretch and force quench for the stretch-quench cycles $13$, $14$, $15$, $16$ and $17$
(middle panel) are denoted by solid green and dash.dotted blue arrows.

\bigskip

{\bf Figure 5.} Typical unfolding-refolding trajectories of $X/a$ (black) and $Q$ (red) for $S1$ as functions 
of time $t$, simulated by applying four stretch-quench cycles at the pulling force $f_{S}$$=$$40 pN$ and 
quenched force $f_Q$$=$$0$. The duration of relaxation time $T$$=$$102ns$.

\bigskip

{\bf Figure 6.} Examples of unfolding-refolding trajectories of $X/a$ (black) and $Q$ (red) for $S1$ as a 
function of time $t$. The pulling force is $f_{S}$$=$$40 pN$ and the quenched force is $f_Q$$=$$0$. The duration 
of relaxation time $T$$=$$150ns$.

\bigskip

{\bf Figure 7.} Same as Figure 6 except $T$$=$$240ns$.

\bigskip

{\bf Figure 8.} Histograms of forced unfolding times $P(t)$ and the joint distributions of unfolding times 
separated by relaxation periods of the quenched force $P(T,t)$. The distribution functions are constructed from 
single unfolding-refolding trajectories of $S1$ simulated in stretch-quench cycles of $f_S$$=$$80pN$ and 
$f_Q$$=$$0$ for $T$$=$$15 ns$, $48 ns$ and $86 ns$. Simulated distributions are shown by red bars with the 
contribution to global unfolding events from coiled conformations $\{ C \}$ indicated by an arrow for 
$T$$=$$86 ns$. The results of the numerical fits obtained by using Eqs. (\ref{2.1})-(\ref{2.5}) 
are represented by solid lines. The energy landscape parameters of $S1$ are summarized in the Table.

\bigskip

{\bf Figure 9.} Histograms of forced unfolding times $P(t)$ and $P(T,t)$ constructed from single 
unfolding-refolding trajectories for $S1$. The stretch-quench cycles were simulated with $f_S$$=$$40pN$ and 
$f_Q$$=$$0$ for $T$$=$$24 ns$, $54 ns$ and $102 ns$. Simulated distributions are shown by red bars with the 
contribution to global unfolding events from coiled conformations $\{ C \}$ indicated by an arrow for
$T$$=$$102 ns$. The results of numerical fit obtained by using Eqs. (\ref{2.1})-(\ref{2.5}) are represented by 
solid lines. The values of the parameters are given in the Table.

\end{document}